\documentclass{article}

\usepackage{emulateapj,apjfonts}

\input epsf

\lefthead{M. Rampp and H.-Th. Janka}
\righthead{CORE COLLAPSE AND POST-BOUNCE EVOLUTION FOR 15 M$_{\odot}$ STAR}

\begin{document}

\title{SPHERICALLY SYMMETRIC SIMULATION WITH BOLTZMANN NEUTRINO TRANSPORT
OF CORE COLLAPSE AND POST-BOUNCE EVOLUTION OF A 15 M$_{\odot}$ STAR}

\author{Markus Rampp\altaffilmark{1} and  
        H.-Thomas Janka\altaffilmark{1} }

\altaffiltext{1}{Max-Planck-Institut f\"ur Astrophysik, 
                 Karl-Schwarzschild-Str.\ 1, D-85741 Garching, Germany;
                 mjr@mpa-garching.mpg.de; thj@mpa-garching.mpg.de}

\begin{abstract}
We present a spherically symmetric, Newtonian core-collapse 
simulation of a 15 M$_{\odot}$ star with
a 1.28 M$_{\odot}$ iron core. The time-, energy-, and angle-dependent transport of
electron neutrinos ($\nu_e$) and antineutrinos ($\bar\nu_e$)
was treated with a new code which iteratively solves the Boltzmann equation
and the equations for neutrino number, energy and momentum to order $O(v/c)$
in the velocity $v$ of the stellar medium.
The supernova shock expands to a maximum radius of 350$\,$km 
instead of only $\sim 240\,$km as in a comparable calculation with 
multi-group flux-limited diffusion (MGFLD) by Bruenn, Mezzacappa, \& Dineva (1995).
This may be explained by stronger neutrino heating due to the more accurate transport
in our model. Nevertheless, after 180$\,$ms of expansion
the shock finally recedes to a radius around 250$\,$km 
(compared to $\sim 170\,$km in the MGFLD run). The effect of an accurate 
neutrino transport is helpful, but not large enough to cause an 
explosion of the considered 15 M$_{\odot}$ star.
Therefore postshock convection and/or an enhancement of the core neutrino luminosity
by convection or reduced neutrino opacities in the neutron star seem necessary for
neutrino-driven explosions of such stars. We find an electron fraction 
$Y_e > 0.5$ in the neutrino-heated matter, which suggests that the
overproduction problem of neutron-rich nuclei with mass numbers $A\approx 90$
in exploding models may be absent when a Boltzmann solver is used for
the $\nu_e$ and $\bar\nu_e$ transport.
\end{abstract}


\section{INTRODUCTION}

The mechanism of supernova explosions of massive stars is still not satisfactorily
understood. Detailed numerical models showed
that the hydrodynamic shock, which is launched when the collapsing
stellar core bounces abruptly by the stiffening of the equation of state (EoS)
at nuclear densities, cannot propagate out promptly but stalls because
of energy losses due to photodisintegration of iron-group nuclei and
neutrino emission from the shock-heated matter (e.g., Bruenn 1985, 1989a,b;
Myra et al.~1987). Early suggestions that energy deposition by neutrinos
might cause an explosion reach back to 
Colgate \& White (1966). The modern version of the neutrino-driven ``delayed''
explosion mechanism is due to Wilson (1985), who found that neutrino
energy deposition can revive the stalled shock on a time scale of several hundred 
milliseconds after bounce (Bethe \& Wilson 1985). Because of the complexity
of the involved physics and the low efficiency of the neutrino energy transfer it
remained unclear for years whether the explosions are sufficiently energetic
and whether the delayed mechanism works for a larger range of stellar masses
(Wilson et al.\ 1986; Bruenn 1993). Recognizing that neutron-finger convection 
in the newly formed neutron star could increase the neutrino luminosities, 
Wilson \& Mayle (1988, 1993) managed to obtain healthy explosions.
However, the question of neutron star convection is not finally settled and
currently it is not clear whether neutron-finger instabilities or Ledoux convection
(Burrows 1987, Pons et al.~1999) or quasi-Ledoux convection (Keil, Janka, \& M\"uller 1996;
Janka \& Keil 1998) or none (Bruenn et al.~1995; Bruenn \& Dineva 1996;
Mezzacappa et al.~1998a) occur and how they affect the explosion.

Multi-dimensional hydrodynamic models 
(Herant et al.\ 1994; 
Miller, Wilson, \& Mayle 1993; Burrows, Hayes, \& Fryxell 1995; 
Janka \& M\"uller 1996;
Mezzacappa et al.\ 1998b) have demonstrated the existence and
the importance of convective overturn in the neutrino-heating layer behind the 
supernova shock. Driven by a negative entropy gradient which emerges behind the
weakening prompt shock and is enhanced by the neutrino energy deposition, the
convective motions transport energy from the region of strongest
heating to the shock, thus raising the postshock pressure and pushing
the shock farther out. At the same time, cold, low-entropy matter is advected downward
where it can readily absorb energy from the upstreaming neutrinos. These 
hydrodynamic instabilities have a bearing on the measured kick velocities of 
pulsars (Lyne \& Lorimer 1994, Cordes \& Chernoff 1998)
and the anisotropies observed in many supernovae. They are essential to
understand the production of radioactive elements in the vicinity of the nascent
neutron star and their large-scale mixing into the hydrogen and helium layers of 
the exploding star (Kifonidis et al.\ 2000).

All multi-dimensional simulations have so far been carried out with serious
simplifications of the neutrino transport. Even the most
advanced spherically symmetric post-bounce models have only employed MGFLD
(Bruenn 1993, Bruenn et al.\ 1995) until recently. The significance of 
an accurate neutrino transport for the delayed explosion
mechanism, however, has long been recognized (Janka 1991, Messer et al.\ 1998, 
Yamada, Janka, \& Suzuki 1999, Burrows et al.\ 2000). It is therefore a natural 
step that a new generation of supernova models will employ schemes
based on a solution of the Boltzmann equation.
In fact, Mezzacappa et al.~(2000) have published results for
a 13 M$_{\odot}$ star which show that a better
transport can make a qualitative change to the outcome of the simulations.
However, they considered a model with an exceptionally small iron core of
1.17 M$_{\odot}$ (Nomoto \& Hashimoto 1988) and the explosion energy was
only $0.41 \times 10^{51}\,$erg at a post-bounce time
of $\sim 550\,$ms. The growth rate of this energy of $0.05\times 10^{51}\,$erg
per 100$\,$ms cannot easily be extrapolated in time and will probably not increase 
the explosion energy significantly, because the density around the mass cut drops
rapidly and the heating region is evacuated by the developing bifurcation 
between neutron star and ejecta. 

In this {\em Letter} we present results for a Newtonian simulation of a 15 
M$_{\odot}$ star with a 1.28 M$_{\odot}$ iron core (Woosley \& Weaver 1995)
which show that an accurate neutrino transport does not produce an explosion
for this star in spherical symmetry.

\section{NUMERICAL METHODS}

We have developed a new transport code which determines the neutrino phase-space 
distribution by iteratively solving the radiation moment equations
for neutrino energy and momentum coupled to the Boltzmann equation.
The code takes into account effects due to the motion of the stellar medium 
to order $O(v/c)$ and determines the neutrino quantities in a comoving frame
of reference (Rampp 2000; Rampp \& Janka 2000, in preparation). 
It allows a general relativistic treatment, but for comparison with 
published results we have restricted ourselves to the Newtonian case.
The angle dependence of the distribution function is
accounted for by the use of a grid of tangent rays which exploits spherical
symmetry. Closure of the set of moment equations is achieved by a variable
Eddington factor calculated from the solution of the Boltzmann equation, and
the integro-differential character of the latter is tamed by making use of the 
integral moments of the neutrino distribution as obtained from the moment
equations. The method is similar to the one described by Burrows et al.~(2000).
In order to fulfill lepton number conservation, we employ additional moment 
equations for neutrino number density and number flux. Severe time step 
restrictions are avoided and proper establishment of equilibrium is ensured
by integrating the set of transport equations implicitly in time.
The stiff character of the source terms for neutrino energy and lepton number 
requires a simultaneous implicit update of the temperature and
electron fraction of the stellar medium. 

The transport is coupled to the hydrodynamics code {\em Pro\-me\-theus},
which integrates the continuity equations for mass, momentum, energy and particle
species in a conservative way on a moving radial grid by explicit time stepping.
The integration is accurate to second order in space and time.
Shocks are treated as local Riemann problems at
the zone interfaces (Fryxell, M\"uller \& Arnett 1989). The source terms for
energy and momentum due to gravity and neutrinos, and for lepton number due to
neutrino emission/absorption are handled by an operator-splitting technique.
The stellar background and the neutrinos are evolved on different 
radial grids and with different time steps, which are constrained by changes
per transport step (which is typically larger than the hydrodynamical step)
of at most 10\% for the neutrino quantities and 5\% for the
fluid quantities. Interpolation between both grids is done in a conservative 
manner.

We used the EoS of Lattimer \& Swesty (1991) (with nuclear incompressibility 
modulus of $K = 180$ MeV) which is extended to
densities and temperatures below the regime of nuclear statistical equilibrium
by an ideal gas equation of state, corrected for Coulomb-lattice effects, 
that includes arbitrarily relativistic and degenerate electrons and positrons,
photons, and a mixture of predefined nuclear species.
Nuclear burning was not taken into account in the present simulation.

The hydrodynamics was solved on a grid with 400 radial zones out to 20000$\,$km,
which were moved with the matter of the iron core during collapse
to ensure good spatial resolution at all times, and kept fixed later. 
For the transport we used a Eulerian grid with 210 geometrically spaced 
radial zones, 230 tangent rays and 27 energy bins geometrically distributed between
0 and 380 MeV, the zone center of the first zone being at 1 MeV. The quality of
the energy conservation limits the error in the net energy deposition by neutrinos to 
$< 5\times 10^{49}\,$erg, and lepton number is globally conserved to better than
0.1\%.

The present simulation includes only $\nu_e$ and $\bar\nu_e$. The corresponding 
rates for charged-current and neutral-current reactions with nucleons and nuclei 
and for neutrino-electron scattering  
were taken from Bruenn (1985), Mezzacappa \& Bruenn (1993) and Bruenn \&
Mezzacappa (1997). We neglect production and annihilation of $\nu_e\bar\nu_e$
pairs, which are of minor importance compared to the charged-current reactions
with nucleons. A detailed comparison of core-collapse results with
published models of Bruenn \& Mezzacappa (1997) showed excellent agreement.
Disregarding muon and tau neutrinos and antineutrinos and $\nu_e\bar\nu_e$ pair 
processes has virtually no effect on the neutrino heating (see Bruenn 1993).

\section{RESULTS}

\begin{figure}
\plotone{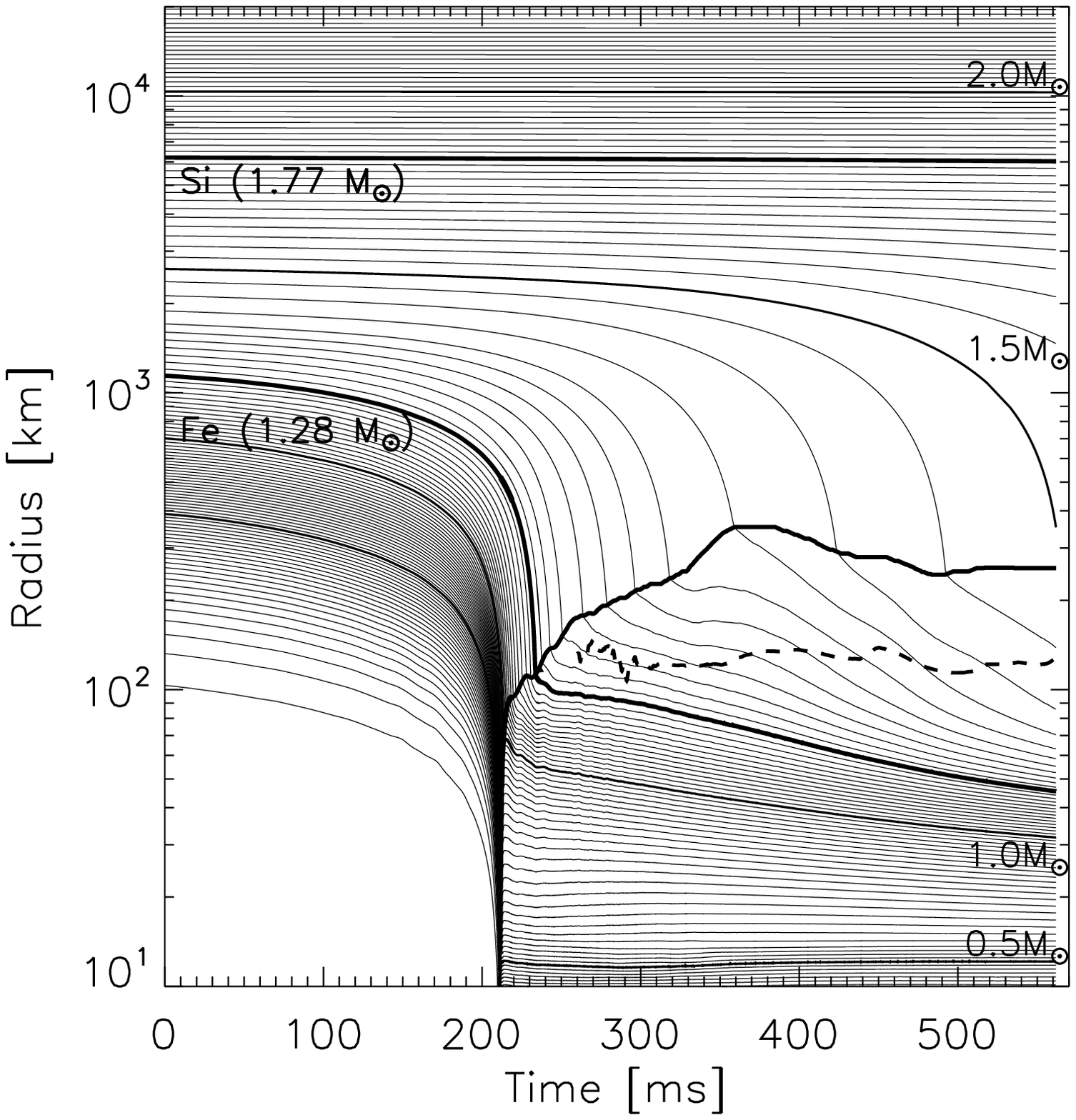}
\caption{
Trajectories of selected mass shells vs.\ time from the start of the simulation.
The shells are equidistantly spaced in steps of 0.02 M$_{\odot}$, and the
trajectories of the outer boundaries of the iron core (at 1.28 M$_{\odot}$)
and of the silicon shell (at 1.77 M$_{\odot}$) are indicated by bold lines.
The shock is formed at 211$\,$ms. Its position is also marked by a bold line.
The dashed curve shows the position of the gain radius.}
\label{fig:1}
\end{figure}

\begin{figure}
\plotone{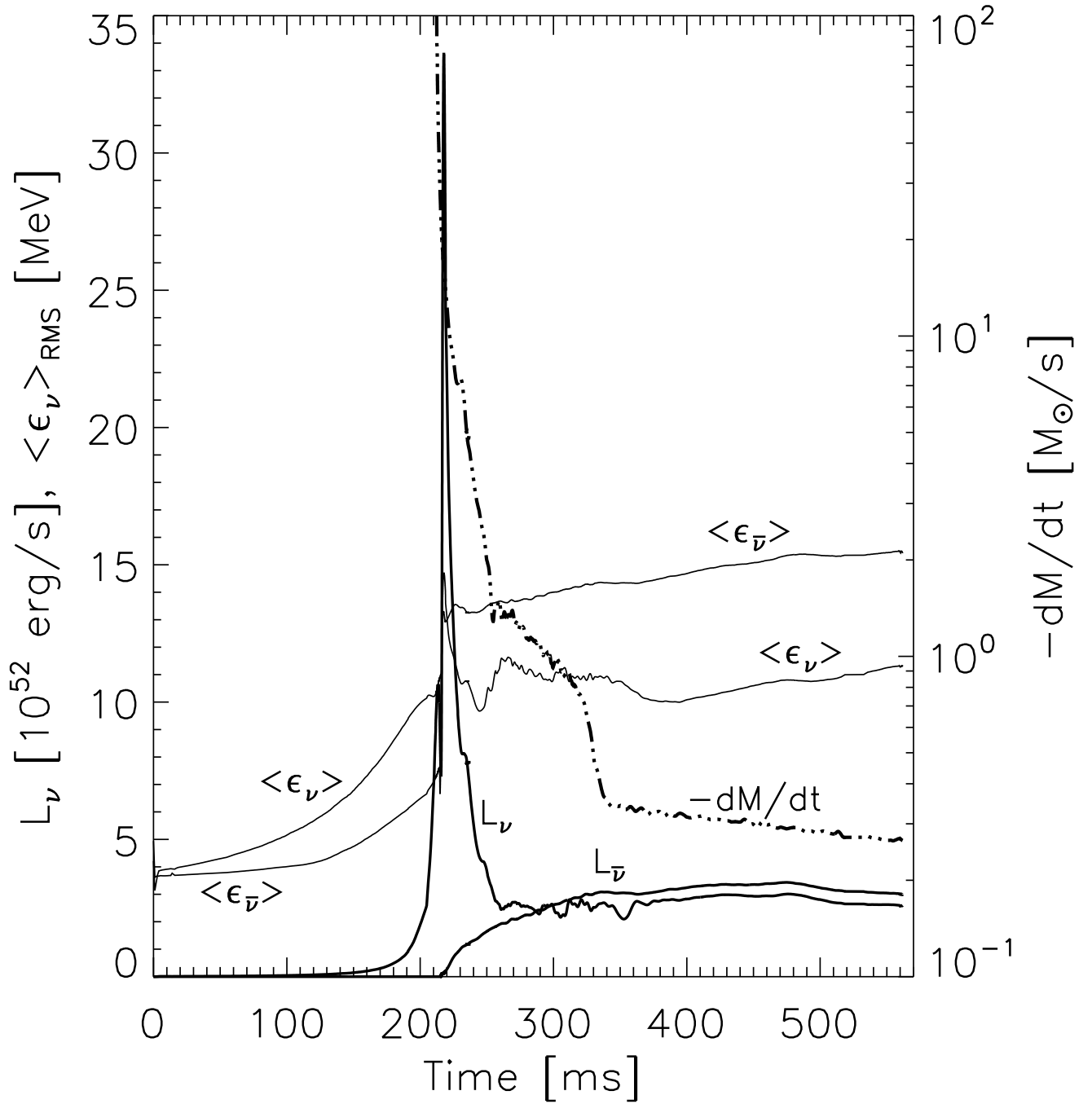}
\caption{
Comoving frame electron neutrino and antineutrino
luminosities (bold solid lines) and rms energies (thin solid lines) at
1000$\,$km as functions of time. Also shown is the mass accretion rate
through the shock (dashed-dotted line, scale on the right ordinate).}
\label{fig:2}
\end{figure}

Figure~\ref{fig:1} shows the trajectories of selected mass shells as a function
of time. The
bounce shock forms 211.6$\,$ms after the onset of the collapse at a radius of 
12.5$\,$km with an enclosed mass of $\sim 0.62$ M$_{\odot}$. The central density
at this time is $\rho_{\mathrm c} = 3.3\times 10^{14}\,$g$\,$cm$^{-3}$
(cf.\ Bruenn \& Mezzacappa 1997). By the rapid accretion of mass (Fig.~\ref{fig:2})
the shock is pushed out to $\sim 240\,$km.
When the accretion rate drops significantly at $\sim 120\,$ms after bounce,
neutrino heating is able to support further shock expansion to a radius of 
350$\,$km. After some time, however, the shock retreats again and finally 
turns into a standing accretion shock around 250$\,$km, still within the collapsing
silicon shell of the progenitor star. No indication for the possibility of an explosion 
was visible when the simulation was terminated at 350$\,$ms after bounce. At this
time the shock was stagnant and enclosed a mass of 1.5 M$_{\odot}$ with increasingly
negative postshock velocities. The decreasing density in the neutrino-heating 
region and the decay of the $\nu_e$ and $\bar\nu_e$ luminosities do not give
hope for a later rejuvenation of the shock.
The overall evolution in our simulation is very similar to
model WPE15ls(180)Newt20 in Bruenn et al.~(1995), who used MGFLD for
the neutrino transport. The most obvious difference
is a larger maximum radius of the shock, 350$\,$km compared to only $\sim 240\,$km
in the calculation by Bruenn et al.~(1995). Also, the shock is able to stay near
its maximum radius for a longer time and afterwards does not recede as far
as in model WPE15ls(180)Newt20. 

The $\nu_e$ and $\bar\nu_e$ luminosities and the root-mean-squared energies
at 1000$\,$km are shown as functions of time in Fig.~\ref{fig:2}. The
prompt $\nu_e$ burst with a peak luminosity of 
$3.36\times 10^{53}\,$erg$\,$s$^{-1}$ arrives at this radius
only $\sim 6\,$ms after core bounce. About 50$\,$ms after bounce the $\nu_e$ and 
$\bar\nu_e$ luminosities have become roughly equal with a fairly stable value of
(2.5--$3)\times 10^{52}\,$erg$\,$s$^{-1}$. By the end of our simulation they begin
to decrease slowly, different from the mean energies, which show a gradual rise
to 11.2 MeV for $\nu_e$ and 15.5 MeV for $\bar\nu_e$.

\begin{figure}
\plotone{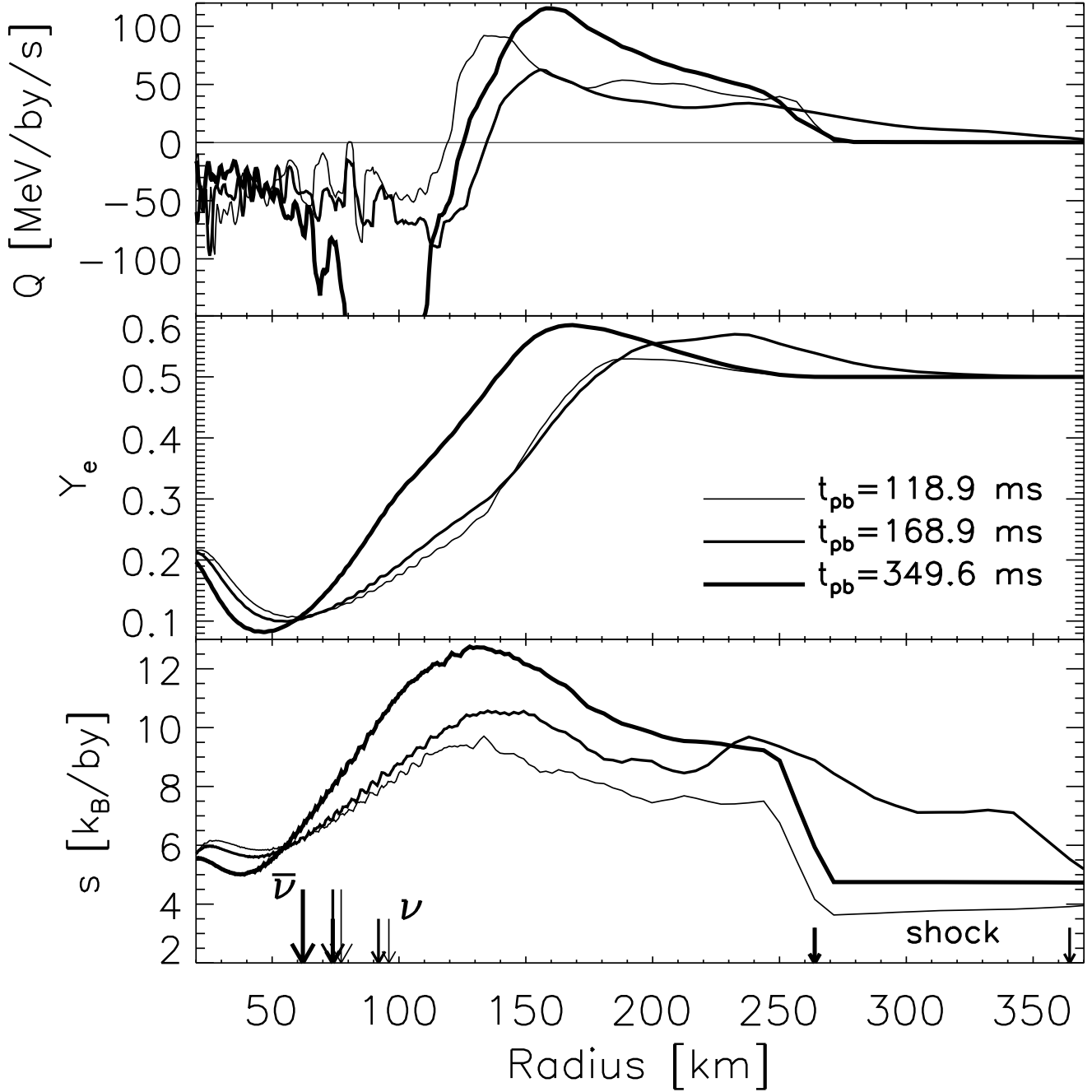}
\caption{
Radial profiles of the net energy deposition rate by $\nu_e$ and $\bar\nu_e$, $Q$,
(top), of the electron fraction $Y_e$ (middle) and of the entropy per baryon, $s$, 
(bottom) at times 119$\,$ms (thin lines), 169$\,$ms (medium lines) and 350$\,$ms
(bold lines) after core bounce. The positions of the shock and of the $\nu_e$-
and $\bar\nu_e$-spheres are indicated.}
\label{fig:3}
\end{figure}

In Fig.~\ref{fig:3} we present profiles of the net energy deposition rate by
$\nu_e$ and $\bar \nu_e$, of the electron fraction $Y_e$ and of the
entropy per baryon for times 330$\,$ms, 380$\,$ms
and 561$\,$ms. The gain radius, below which net neutrino cooling
and above which net heating occurs, is at 120--140$\,$km (see also
Fig.~\ref{fig:1}). The heating rates peak somewhat outside the
gain radius and reach up to $\sim 120$ MeV$\,$s$^{-1}$ per baryon. The cooling 
rate below the gain radius 
can exceed 200 MeV$\,$s$^{-1}$ per baryon at late times. Maximum entropies
around 13 $k_{\mathrm B}$ per nucleon are seen at the end of the
simulation, when the density behind the shock is lowest because of the decreasing
mass accretion rate. The negative entropy gradient 
implies potential instability against convective overturn in the region
between maximum heating and supernova shock. In this layer $Y_e$
climbs to values {\em larger} than 0.5 and also develops a negative gradient.
Values $Y_e > 0.5$ were also found by Mezzacappa et al.~(2000)
in the neutrino-heated ejecta behind the outgoing shock for the successful
explosion of a 13 M$_{\odot}$ star.
The neutronization of the neutrino-heated medium is determined by the absorption
of $\nu_e$ on neutrons and of $\bar\nu_e$ on protons and the inverse processes.
It is sensitive to the luminosities and spectra but also to the
angular distributions of the neutrinos in the heating region, which govern
the efficiency of energy deposition as well as the lepton exchange with the medium. 
Since $\nu_e$ decouple at a larger radius than $\bar\nu_e$, their distribution
is more isotropic in the heating region, leading to a higher probability of 
$\nu_e$ absorption and thus to an increase of $Y_e$. This is enhanced by the
recombination of $\alpha$ particles (Fuller \& Meyer 1995).

\section{CONCLUSIONS}

Our spherically symmetric, Newtonian simulation of a 15 M$_{\odot}$ star with a
1.28 M$_{\odot}$ iron core, using a new Boltzmann solver for the neutrino transport,
did not give an explosion until 350$\,$ms after core bounce, although the
shock reached a larger maximum radius than in a comparable MGFLD simulation of
Bruenn et al.\ (1995). This is probably explained by stronger
neutrino heating of the postshock medium with the more accurate
Boltzmann transport. Since both simulations were done with the same 
progenitor, EoS, and neutrino opacities, and excellent agreement
during the core-collapse phase was found, uncertainties due to the different 
numerics seem to be minimized. Although we have only included $\nu_e$
and $\bar\nu_e$ in our simulation, we consider our conclusions as solid, because
muon and tau neutrinos would drain energy from the $\nu_e$ and $\bar\nu_e$ luminosities
but contribute to the postshock heating only at an insignificant level due to the
lack of charged-current interactions. The main effect of adding pair processes would 
be a weakening of the early shock propagation by additional energy losses. Also general
relativity would probably hamper an explosion (Fryer 1999), but the situation is 
still ambiguous (De Nisco, Bruenn, \& Mezzacappa 1998; Baron 1988).

The importance of an accurate $\nu_e$ and $\bar\nu_e$ transport is
emphasized by the finding that $Y_e > 0.5$ in the region of net neutrino 
energy deposition. This is interesting because $Y_e \la 0.48$
was obtained in the neutrino-heated ejecta in supernova models, e.g., 
by Herant et al.\ (1994), Burrows et al.\ (1995) and Janka \& M\"uller (1996),
causing a large overproduction of neutron-rich nuclei around $N = 50$ and $A\approx 90$
($^{88}$Sr, $^{89}$Y, $^{90}$Zr). This is in conflict with measured galactic 
abundances. With values $Y_e > 0.5$ this problem disappears (Hoffman et al.\ 1996).

Using their Boltzmann solver for the neutrino transport, Mezzacappa et al.\ (2000)
obtained a successful, but weak explosion in case of a 13 M$_{\odot}$
progenitor with an extraordinarily small iron core of 1.17 M$_{\odot}$.
For a 15 M$_{\odot}$ star with a larger core (and therefore most
likely also for more massive progenitors), we cannot confirm a
qualitative difference from spherically symmetric simulations with MGFLD transport,
although we find important quantitative differences with our more 
accurate neutrino transport. In order to obtain explosions via the neutrino-heating
mechanism, multi-dimensional simulations seem indispensable for stars with 
typical iron core masses. Convection inside the neutron star (Keil et al.\ 1996)
or lower neutrino opacities --- due to suppression relative to the standard description
by nucleon correlation effects (e.g., Janka et al.\ 1996, Burrows \& Sawyer 1998,
Reddy et al.\ 1999) --- could raise the
neutrino emission significantly on the relevant time scale of a few 100$\,$ms
after bounce, and convective overturn in the 
postshock region has been shown by several groups to support the explosion.

\acknowledgments
Support by Deutsche Forschungsgemeinschaft grant SFB 375 f\"ur Astro-Teilchenphysik 
is acknowledged. The NEC SX-5/3C of the Rechenzentrum Garching was used for the 
computations.


\clearpage

\clearpage

\end{document}